\documentclass[conference]{IEEEtran}
\IEEEoverridecommandlockouts
\usepackage{cite}
\usepackage{amsmath,amssymb,amsfonts}
\usepackage{algorithmic}
\usepackage{graphicx}
\usepackage{textcomp}
\usepackage{xcolor}
\def\BibTeX{{\rm B\kern-.05em{\sc i\kern-.025em b}\kern-.08em
    T\kern-.1667em\lower.7ex\hbox{E}\kern-.125emX}}

\newcommand{\Der}{\textrm{d}}

\usepackage[pscoord]{eso-pic}
\newcommand{\placetextbox}[3]{
    \setbox0=\hbox{#3}
     \AddToShipoutPictureFG*{
     \put(\LenToUnit{#1\paperwidth},\LenToUnit{#2\paperheight}){
      \vtop{{\null}\makebox[0pt][c]{#3}}}
    }
}
\placetextbox{.2}{0.055}{ 978-3-903176-44-7~\copyright~2022 IFIP}

\begin{document}

\title{Quantum Limits on the Capacity of Multispan Links with Phase-Sensitive Amplification\\
\thanks{This work is a part of the project ``Quantum Optical Technologies'' carried out within the International Research Agendas programme of the Foundation for Polish Science co-financed by the European Union under the European Regional Development Fund.}
}

\author{\IEEEauthorblockN{Karol {\L}ukanowski\IEEEauthorrefmark{1}\IEEEauthorrefmark{2}\IEEEauthorrefmark{4}, Marcin Jarzyna\IEEEauthorrefmark{1}, and Konrad Banaszek\IEEEauthorrefmark{1}\IEEEauthorrefmark{2}}
\IEEEauthorblockA{\IEEEauthorrefmark{1}Centre for Quantum Optical Technologies, CeNT,
University of Warsaw, Banacha 2c, 02-097 Warszawa, Poland \\
\IEEEauthorrefmark{2}Faculty of Physics, University of Warsaw, Pasteura 5, 02-093 Warszawa, Poland\\
\IEEEauthorrefmark{4}E-mail: k.lukanowski@cent.uw.edu.pl}
}

\maketitle

\begin{abstract}
The capacity of a linear attenuating optical channel with the signal regenerated using quantum-limited phase-sensitive amplifiers is analyzed for conventional and generalized detection scenarios.
The continuous model of distributed amplification determines the attainable capacity for long-haul links under the total power constraint.
\end{abstract}


\section{Introduction}

The development of quantum technologies provides a strong driving force to improve the performance of photonic components and devices \cite{PelucchiNatRevPhys2021}. While the principal motivation is to reach performance levels that would enable new functionalities with the flagship example of quantum key distribution \cite{PirandolaAOP2020}, these advances may greatly benefit also regular optical communications. It is therefore vital to investigate how the availability of improved photonic components and devices would affect the design and the modelling of optical networks and to quantify the attainable enhancements. The complete analysis needs to be based on the quantum theory of electromagnetic radiation \cite{Shapiro2009} which includes fundamental sources of quantum noise that determine the ultimate performance limits of optical communication systems \cite{BanaszekJLT2020}.

Recent years have witnessed an impressive progress in the technology of phase-sensitive amplifiers \cite{KakandeOFC2011,UmekiOpEx2013,OlssonNCOMM2018}.
The favorable noise figure of phase-sensitive amplification (PSA) makes it a promising approach to increase the transmission rates in long-haul optical communication systems.
The purpose of this contribution is to analyze theoretically the capacity limits of a multispan point-to-point link with PSA regeneration taking fully into account quantum mechanical fluctuations of the propagating optical field \cite{GerryKnight2004}. It is shown that these fluctuations introduce a fundamental noise penalty that causes a decline in the capacity of the resulting communication channel even for ideal quantum-limited PSA with a 0~dB noise figure. The analysis is carried out under the total power constraint for the propagating field and assumes one of two detection scenarios at the channel output \cite{BanaszekJLT2020}: either a shot-noise-limited (SNL) single-quadrature measurement, or the most general optimized measurement strategy permitted by quantum mechanics. The efficiency of the latter is characterized by the Gordon-Holevo capacity limit adapted to phase-sensitive Gaussian channels \cite{SchaeferPRL2013,SchaeferXXX2016}.
An approximate analytical solution is derived in the continuous limit of distributed regeneration which facilitates comparison with the phase-insensitive amplification (PIA) scenario discussed previously \cite{JarzynaECOC2019}.

\section{Channel model}

We start by presenting the theoretical model used in numerical simulations of a multispan optical communication link.
Let $x^I$ and $x^Q$ denote two conjugate $I$ and $Q$ quadratures of a single mode of the electromagnetic field. It will be convenient to decompose
$x^I = x_S^I + x_N^I$ and $x^Q = x_S^Q + x_N^Q$,
where $x_S^I $ and $x_S^Q$ are the mean values that carry the modulated symbols, while zero-mean variables $x_N^I$ and $x_N^Q$ describe quantum fluctuations. Units are chosen such that vacuum field fluctuations correspond to the variance
$\textrm{Var}  [x_N^I] = \textrm{Var}  [x_N^Q] = 1/2$. Formally, the variables $x^I$ and $x^Q$ defined above parameterize the quantum mechanical Wigner function of the electromagnetic field mode under consideration \cite{GerryKnight2004}.
For Gaussian signal modulation and Gaussian noise it is helpful to introduce the signal power and the noise power in individual quadratures:
\begin{equation}
    \begin{IEEEeqnarraybox}[][c]{rCl"rCl}
        S^I &=& \textrm{Var}  [x_S^I], & N^I  &=& \textrm{Var}  [x_N^I],\\
        S^Q &=& \textrm{Var}  [x_S^Q], & N^Q  &=& \textrm{Var}  [x_N^Q].
    \end{IEEEeqnarraybox}
\end{equation}
The Heisenberg uncertainty relation requires that $N^I \cdot N^Q \ge 1/4$.
The total energy per mode expressed as the mean photon number $\bar{n}$ is given by
\begin{equation}
\bar{n} =  (S^I + S^Q + N^I + N^Q)/2 -1/2.
\label{Eq:nbar}
\end{equation}
The negative $-1/2$ term is needed to subtract vacuum fluctuations when converting the quadrature variance budget to the optical energy carried by the mode.

\begin{figure*}
\begin{center}
\includegraphics[width=0.95\textwidth]{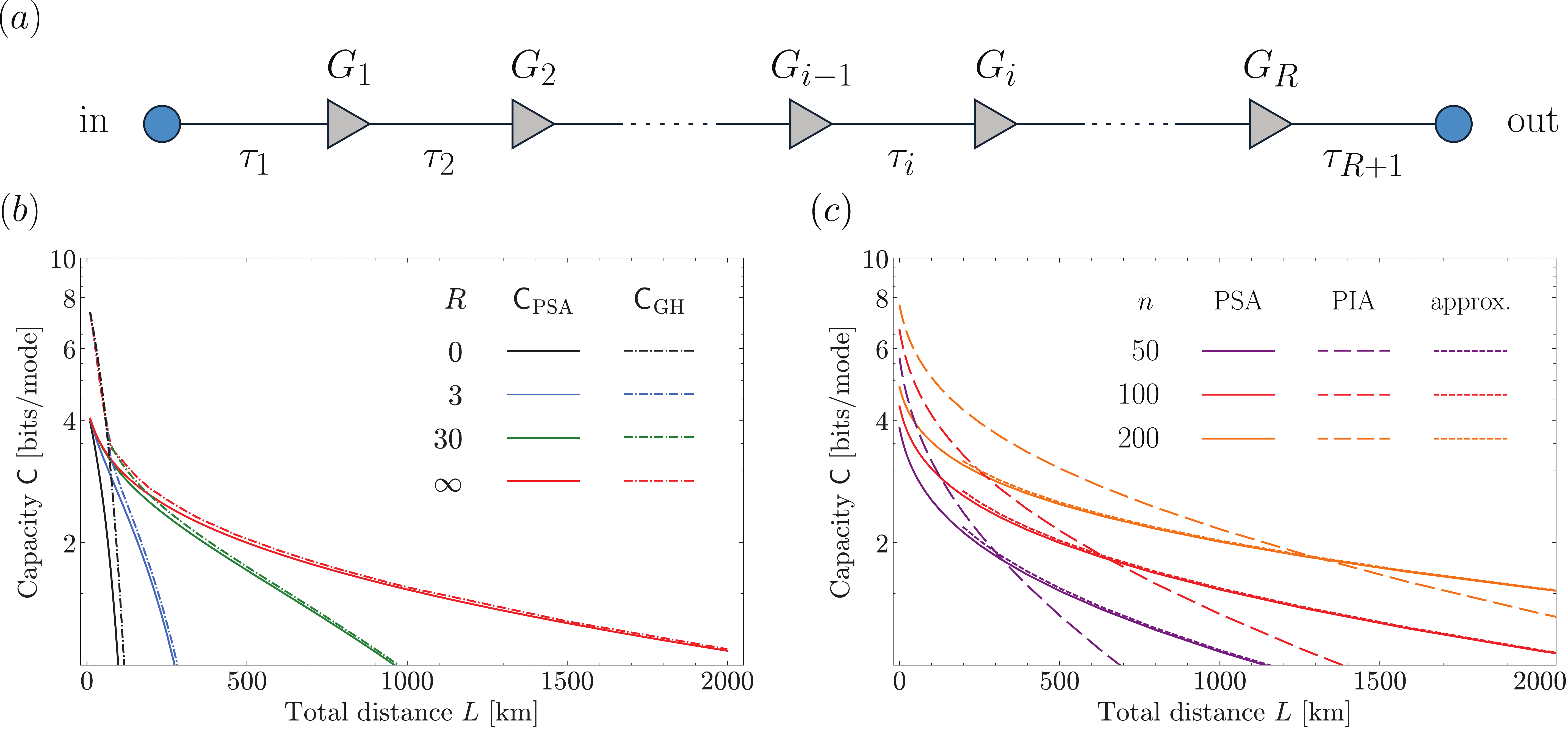}
\end{center}
\caption{(a) A multispan link with $R+1$ segments with transmissions $\tau_i$, $i=1,\ldots,R+1$. The signal is regenerated using $R$ phase-sensitive amplifiers with gains $G_i$. (b) Channel capacity as a function of distance for $\bar{n}=100$ input mean photon number optimized over the locations and gains of $R$ phase-sensitive amplifiers under the total power constraint. Solid lines depict the conventional SNL single-quadrature detection scenario, whereas dashed-dotted lines correspond to the generalized scenario described by the Gordon-Holevo (GH) expression adapted to phase-sensitive Gaussian channels. The case of distributed amplification is labeled as $R=\infty$. The standard value of $0.2~\textrm{dB}/\textrm{km}$ for channel attenuation has been used. (c) Conventional-detection Shannon capacity as a function of distance for distributed amplification and the input mean photon number $\bar{n} = 50$ (purple), $100$ (red), and $200$ (orange). The numerical solutions of (\ref{Eq:DistributedDiff}) describing the PSA scenario (solid lines) are compared with the approximation (\ref{Eq:CPSA}) (dotted lines) and the PIA regeneration scenario analyzed in \cite{JarzynaECOC2019} (dashed lines).}
\label{Fig:PSA}
\end{figure*}

Consider now a link shown in Fig.~\ref{Fig:PSA}(a) in the form of $R+1$ attenuating spans with respective power transmissions $\tau_i$, $i=1,2,\ldots, R+1$ and the total length $L$. The signal is regenerated using $R$ phase-sensitive amplifiers with gains $G_i$.  For quantum-limited PSA the recursive relation for the optical field parameters over the $i$th span followed by an amplifier reads
\begin{equation}
    \begin{IEEEeqnarraybox}[][c]{rCl'rCl}
        S_{i}^I &=& G_{i} \tau_{i} S_{i-1}^I, &  N_{i}^I &=& G_{i} [\tau_{i} N_{i-1}^I + (1-\tau_{i})/2 ],\\[3pt]
        S_{i}^Q &=& G_{i}^{-1} \tau_{i} S_{i-1}^Q, & N_{i}^Q &=& G_{i}^{-1} [\tau_{i} N_{i-1}^Q + (1-\tau_{i})/2 ].
    \end{IEEEeqnarraybox}
\label{Eq:Recursion}
\end{equation}
where $i=0$ denotes the channel input and $i=R+1$ specifies the channel output. Equations~(\ref{Eq:Recursion}) can be used also for the last unamplified span by setting $G_{R+1} = 1$. Importantly, the terms $(1-\tau_{i})/2$ in the expressions for $N_{i}^I$ and $N_{i}^Q$ stem from vacuum fluctuations that are added in the course of propagation through an attenuating medium  \cite{GerryKnight2004}.
In the following indices `in' for $i=0$ and `out' for $i=R+1$ will be used for clarity.

\section{Numerical optimization}

The capacity of the link described above has been optimized with respect to the locations $l_i$ and gains $G_i$ of the amplifiers under the total power constraint, i.e.,\ that at no point along the link the right hand side of (\ref{Eq:nbar}) exceeds the input mean photon number $\bar{n}$. Two communication scenarios have been considered. In the conventional scenario, information is encoded only in the $I$ quadrature of ideal laser light with $N^I_\textrm{in} = N^Q_\textrm{in} =1/2$ at the input, and SNL detection of that quadrature is implemented at the output. The capacity is then given by the standard Shannon expression ${\sf C}_{\textrm{PSA}} = \frac{1}{2} \log_2 (1 + S^I_{\textrm{out}}/N^I_{\textrm{out}})$.
The second, generalized scenario permits an arbitrary distribution of the input power between the $I$ and $Q$ quadratures at the input and allows using non-classical squeezed states \cite{GerryKnight2004} for which either $N^I_\textrm{in}$ or $N^Q_\textrm{in}$ is reduced below $1/2$.
Furthermore, the generalized scenario assumes optimization over all measurements of the optical field at the channel output that are permitted by quantum mechanics. In such a case the capacity ${\sf C}_{\textrm{GH}}$ is given by the Gordon-Holevo (GH) formula extended  to the case of phase-sensitive Gaussian channels \cite{SchaeferPRL2013,SchaeferXXX2016} that need to be considered here.

The results of optimizing the channel capacity with PSA regeneration for the input mean photon number $\bar{n}=100$ are shown in Fig.~\ref{Fig:PSA}(b). The standard $0.2~\textrm{dB}/\textrm{km}$ figure for the attenuation has been taken to calculate transmission of individual spans $\tau_i = \exp[-\alpha(l_i-l_{i-1})]$. It is seen that for very short distances the generalized strategy offers advantage which however virtually does not depend on the ability to regenerate the signal. The observed advantage effectively stems from using both field quadratures to transmit information over a loss-only channel \cite{BanaszekJLT2020}. For longer distances PSA becomes useful to boost the link capacity. In this regime the difference between the capacities attainable in the conventional and in the generalized scenarios becomes negligible.
Overall, the capacity is seen to decline with the distance despite assuming ideal, quantum-limited PSA in (\ref{Eq:Recursion}). This effect can be attributed to the amplification of vacuum fluctuations added in the course of propagation through an attenuating medium.

\section{Distributed amplification}

The minimum decline rate for the channel capacity under PSA can be found by considering densely spaced amplifiers at equidistant locations $l_i=i\Delta l$. One can then approximate $\tau_i \approx 1 - \alpha \Delta l$, where $\alpha$ is the attenuation of the propagation medium, and take $G_i \approx 1 + \gamma(l_i) \Delta l$. In the continuous limit of distributed amplification $\Delta l \rightarrow 0$ this gives a set of differential equations
\begin{equation}
        \begin{IEEEeqnarraybox}[
            ][c]{rCl}
                \frac{\Der S^I}{\Der l} &=& [\gamma(l)-\alpha] S^I, \\
                \frac{\Der N^I}{\Der l} &=& [\gamma(l)-\alpha] N^I + \alpha/2, \\
                \frac{\Der S^Q}{\Der l} &=& [-\gamma(l)-\alpha] S^Q, \\
                \frac{\Der N^Q}{\Der l} &=& [-\gamma(l)-\alpha] N^Q + \alpha/2.
        \end{IEEEeqnarraybox}
    \label{Eq:DistributedDiff}
\end{equation}
In the conventional communication scenario, when only the $I$ quadrature is modulated one has $S^Q=0$ and can approximate
$N^Q = 1/2$. The simplified total power constraint ${\Der}(S^I+N^I)/\Der l=0$ yields the amplification factor equal to $\gamma = \alpha[1-(4\bar{n}+1)^{-1}]$. It is seen that full compensation of signal losses is not possible due to the buildup of excess noise through amplification of vacuum fluctuations. The resulting simplified solution for the $I$ quadrature at the channel output reads
\begin{equation}
        \begin{IEEEeqnarraybox}[
            ][c]{rCl}
                S^{I}(L) & = & 2\bar{n} \exp \left( - \frac{\alpha L }{4\bar{n}+1} \right), \\
                N^{I}(L) & = &  2\bar{n} \left[1- \exp \left( - \frac{\alpha L }{4\bar{n}+1} \right) \right] + \frac{1}{2}.
        \end{IEEEeqnarraybox}
\end{equation}
If the vacuum fluctuations level $1/2$ can be neglected compared to $\alpha L$ and $\bar{n}$, the conventional-detection one-quadrature capacity is well approximated by
\begin{IEEEeqnarray}{rCl}
{\sf C}_{\textrm{PSA}} &=& \frac{1}{2} \log_2 \left( 1+ \frac{S^{I}(L)}{ N^{I}(L)}\right) \nonumber \\
&\approx& - \frac{1}{2} \log_2 \left(1-\exp\left(-\frac{\alpha L}{4\bar{n}}\right)\right).
\label{Eq:CPSA}
\end{IEEEeqnarray}
As seen in Fig.~\ref{Fig:PSA}(c), the capacity calculated using the exact solution of the set of differential equations (\ref{Eq:DistributedDiff}) is matched well by the approximate expression derived in (\ref{Eq:CPSA}). It is instructive to compare this result with the case of phase-insensitive amplification (PIA) analyzed in \cite{JarzynaECOC2019}. The dashed curves in Fig.~\ref{Fig:PSA}(c) represent the two-quadrature capacity ${\sf C}_{\textrm{PIA}}$ under conventional SNL coherent detection, which analogously to (\ref{Eq:CPSA}) can be approximated by
\begin{equation}
{\sf C}_{\textrm{PIA}}
\approx - \log_2\left(1-\exp\left(-\frac{\alpha L}{\bar{n}}\right)\right).
\end{equation}
It is seen that while for short distances the benefit of using both quadratures to encode information prevails, PSA becomes advantageous for long-haul links. A comparison of the approximate expressions for ${\sf C}_{\textrm{PSA}}$ and  ${\sf C}_{\textrm{PIA}}$ quantifies the advantage as a four times lower decline rate of the capacity with the distance.
This effect can be attributed to the fact that whereas the performance of a quantum-limited PSA link is degraded only by amplification of vacuum fluctuations contributed by the process of signal attenuation, PIA inevitably adds excess noise to both field quadratures in the amount corresponding to at least 3~dB noise figure.

\section{Conclusions}

The capacity of a multispan link with PSA regeneration has been studied under the total power constraint taking fully into account quantum mechanical fluctuations of the propagating optical field. It has been found that for long-haul links unconventional detection strategies described by the Gordon-Holevo capacity expression do not offer a meaningful benefit compared to conventional SNL single-quadrature detection. A simplified analysis of the continuous limit of distributed amplification indicates that the ultimate advantage of PSA regeneration is four times slower decline rate of the capacity with the distance compared to the PIA scenario.

\section*{Acknowledgment}

Insightful discussions with P. Andrekson, M. Karlsson, J. Schr\"{o}der, and J. P. Turkiewicz are gratefully acknowledged.

\bibliographystyle{IEEEtran}
\bibliography{psabib}

\begin{thebibliography}{10}
\providecommand{\url}[1]{#1}
\csname url@samestyle\endcsname
\providecommand{\newblock}{\relax}
\providecommand{\bibinfo}[2]{#2}
\providecommand{\BIBentrySTDinterwordspacing}{\spaceskip=0pt\relax}
\providecommand{\BIBentryALTinterwordstretchfactor}{4}
\providecommand{\BIBentryALTinterwordspacing}{\spaceskip=\fontdimen2\font plus
\BIBentryALTinterwordstretchfactor\fontdimen3\font minus
  \fontdimen4\font\relax}
\providecommand{\BIBforeignlanguage}[2]{{%
\expandafter\ifx\csname l@#1\endcsname\relax
\typeout{** WARNING: IEEEtran.bst: No hyphenation pattern has been}%
\typeout{** loaded for the language `#1'. Using the pattern for}%
\typeout{** the default language instead.}%
\else
\language=\csname l@#1\endcsname
\fi
#2}}
\providecommand{\BIBdecl}{\relax}
\BIBdecl

\bibitem{PelucchiNatRevPhys2021}
\BIBentryALTinterwordspacing
E.~Pelucchi, G.~Fagas, I.~Aharonovich, D.~Englund, E.~Figueroa, Q.~Gong,
  H.~Hannes, J.~Liu, C.-Y. Lu, N.~Matsuda, J.-W. Pan, F.~Schreck, F.~Sciarrino,
  C.~Silberhorn, J.~Wang, and K.~D. J\"{o}ns, ``The potential and global
  outlook of integrated photonics for quantum technologies,'' \emph{Nature
  Reviews Physics}, 2021. [Online]. Available:
  \url{https://doi.org/10.1038/s42254-021-00398-z}
\BIBentrySTDinterwordspacing

\bibitem{PirandolaAOP2020}
\BIBentryALTinterwordspacing
S.~Pirandola, U.~L. Andersen, L.~Banchi, M.~Berta, D.~Bunandar, R.~Colbeck,
  D.~Englund, T.~Gehring, C.~Lupo, C.~Ottaviani, J.~L. Pereira, M.~Razavi,
  J.~S. Shaari, M.~Tomamichel, V.~C. Usenko, G.~Vallone, P.~Villoresi, and
  P.~Wallden, ``Advances in quantum cryptography,'' \emph{Adv. Opt. Photon.},
  vol.~12, no.~4, pp. 1012--1236, Dec 2020. [Online]. Available:
  \url{https://doi.org/10.1364/AOP.361502}
\BIBentrySTDinterwordspacing

\bibitem{Shapiro2009}
\BIBentryALTinterwordspacing
J.~H. Shapiro, ``The quantum theory of optical communications,'' \emph{IEEE J.
  Sel. Top. Quant. Electr.}, vol.~15, no.~6, pp. 1547--1569, 2009. [Online].
  Available: \url{https://doi.org/10.1109/JSTQE.2009.2024959}
\BIBentrySTDinterwordspacing

\bibitem{BanaszekJLT2020}
\BIBentryALTinterwordspacing
K.~Banaszek, L.~Kunz, M.~Jachura, and M.~Jarzyna, ``Quantum limits in optical
  communications,'' \emph{J. Lightwave Technol.}, vol.~38, no.~10, pp.
  2741--2754, May 2020. [Online]. Available:
  \url{https://doi.org/10.1109/JLT.2020.2973890}
\BIBentrySTDinterwordspacing

\bibitem{KakandeOFC2011}
\BIBentryALTinterwordspacing
J.~Kakande, A.~Bogris, R.~Slav\'{i}k, F.~Parmigiani, D.~Syvridis, M.~Sk\"{o}ld,
  M.~Westlund, P.~Petropoulos, and D.~J. Richardson, ``{QPSK} phase and
  amplitude regeneration at 56 {G}baud in a novel idler-free non-degenerate
  phase sensitive amplifier,'' in \emph{Optical Fiber Communication
  Conference/National Fiber Optic Engineers Conference 2011}.\hskip 1em plus
  0.5em minus 0.4em\relax Optical Society of America, 2011, p. OMT4. [Online].
  Available: \url{https://doi.org/10.1364/OFC.2011.OMT4}
\BIBentrySTDinterwordspacing

\bibitem{UmekiOpEx2013}
\BIBentryALTinterwordspacing
T.~Umeki, M.~Asobe, H.~Takara, Y.~Miyamoto, and H.~Takenouchi, ``Multi-span
  transmission using phase and amplitude regeneration in {PPLN}-based {PSA},''
  \emph{Opt. Express}, vol.~21, no.~15, pp. 18\,170--18\,177, Jul 2013.
  [Online]. Available: \url{https://doi.org/10.1364/OE.21.018170}
\BIBentrySTDinterwordspacing

\bibitem{OlssonNCOMM2018}
\BIBentryALTinterwordspacing
S.~L. Olsson, H.~Eliasson, E.~Astra, M.~Karlsson, and P.~A. Andrekson,
  ``Long-haul optical transmission link using low-noise phase-sensitive
  amplifiers,'' \emph{Nature Communications}, vol.~9, no.~1, p. 2513, Jun 2018.
  [Online]. Available: \url{https://doi.org/10.1038/s41467-018-04956-5}
\BIBentrySTDinterwordspacing

\bibitem{GerryKnight2004}
C.~Gerry and P.~Knight, \emph{Introductory Quantum Optics}.\hskip 1em plus
  0.5em minus 0.4em\relax Cambridge University Press, 2004.

\bibitem{SchaeferPRL2013}
\BIBentryALTinterwordspacing
J.~Sch\"afer, E.~Karpov, R.~Garc\'{\i}a-Patr\'on, O.~V. Pilyavets, and N.~J.
  Cerf, ``Equivalence relations for the classical capacity of single-mode
  {G}aussian quantum channels,'' \emph{Phys. Rev. Lett.}, vol. 111, p. 030503,
  Jul 2013. [Online]. Available:
  \url{https://doi.org/10.1103/PhysRevLett.111.030503}
\BIBentrySTDinterwordspacing

\bibitem{SchaeferXXX2016}
\BIBentryALTinterwordspacing
J.~Sch\"{a}fer, E.~Karpov, O.~V. Pilyavets, and N.~J. Cerf, ``Classical
  capacity of phase-sensitive {G}aussian quantum channels,''
  \emph{arXiv:1609.04119 [quant-ph]}, 2016. [Online]. Available:
  \url{https://arxiv.org/abs/1609.04119}
\BIBentrySTDinterwordspacing

\bibitem{JarzynaECOC2019}
\BIBentryALTinterwordspacing
M.~Jarzyna, R.~Garc{\'\i}a-Patr{\'o}n, and K.~Banaszek, ``Ultimate capacity
  limit of a multi-span link with phase-insensitive amplification,'' in
  \emph{45th European Conference on Optical Communication}, Dublin, Ireland,
  Sep. 2019. [Online]. Available: \url{https://doi.org/10.1049/cp.2019.0742}
\BIBentrySTDinterwordspacing

\end{thebibliography}

\end{document}